# Winning Models for GPA, Grit, and Layoff in the Fragile Families Challenge

Daniel E Rigobon, Eaman Jahani, Yoshihiko Suhara, Khaled AlGhoneim
Abdulaziz Alghunaim, Alex 'Sandy' Pentland, Abdullah Almaatouq[*]

Massachusetts Institute of Technology

[*] To whom correspondence should be addressed; E-mail:
amaatouq@mit.edu

In this paper, we discuss and analyze our approach to the Fragile Families Challenge. The data consisted of over 12,000 features (covariates) about the children and their parents, schools, and overall environments from birth to age 9. Our modular and collaborative approach parallelized prediction tasks, and relied primarily on existing data science techniques, including: (1) data preprocessing: elimination of low variance features, imputation of missing data, and construction of composite features; (2) feature selection through univariate mutual information and extraction of non-zero LASSO coefficients; (3) three machine learning models: Random Forest, Elastic Net, and Gradient-Boosted Trees; and finally (4) prediction aggregation according to performance. The top-performing submissions produced winning out-of-sample predictions for three outcomes: GPA, grit, and layoff. However, predictions were at most 20% better than a baseline that predicted the mean value of the training data for each outcome.

# Acknowledgments

Funding for the Fragile Families and Child Wellbeing Study was provided by the Eunice Kennedy Shriver National Institute of Child Health and Human Development through grants R01HD36916, R01HD39135, and R01HD40421 and by a consortium of private foundations, including the Robert Wood Johnson Foundation. Funding for the Fragile Families Challenge was provided by the Russell Sage Foundation.

The results in this paper were created with software using: Python 3.6.1 (Python Software Foundation; 2017) with packages numpy 1.12.1 (NumPy Developers; 2017), scipy 0.19.0 (SciPy Developers; 2017), matplotlib 2.0.2 (Hunter, Droettboom; 2017), seaborn 0.8.1 (Waskom; 2017), pandas 0.20.1 (The PyData Development Team; 2017), scikit_learn 0.18.1 (Pedregosa et al.; 2011), statsmodels 0.8.0 (Seabold, Skipper, Perktold; 2010), astropy 1.3.2 (Aldcroft, Cruz, Robitaille, Tollerud; 2017), XGBoost 0.6 (Chen, Tianqi, Guestrin; 2016); R 3.4.3 (R Core Team, 2017) with packages data.table 1.10.4-2 (Dowle, Srinivasan, Gorecki, Short, Lianoglou, Antonyan; 2017), Amelia 1.6.2 (Honaker, King, Blackwell; 2012).

# 1. Introduction

In this paper, we describe our individual and team submissions that collectively won first place in three categories in the Fragile Families Challenge (FFC). The Challenge was based on the Fragile Families and Child Wellbeing Study (FFCWS) (Jane Waldfogel, Waldfogel, Craigie, & Brooks-Gunn, 2010; McLanahan & Garfinkel, 2000), which followed thousands of American households for over 15 years and collected information about the children and their parents, schools, and environments. Within this data, six key outcomes were identified: (1) grade point average (GPA) and (2) grit of the child; (3) material hardship and (4) eviction of the household; and (5) layoff and (6) job training of the primary caregiver. Given these outcomes for a small portion of households as training data and approximately 12,000 features[1] from birth to age nine for all households, Challenge participants were tasked with predicting the outcomes for all households. Our best performing submissions were ranked 1st in predicting GPA, grit, and layoff, along with 3rd for job training, 8th for material hardship, and 11th for eviction.

The FFCWS data (Jane Waldfogel et al., 2010; McLanahan & Garfinkel, 2000) has been used in studies attempting to understand causal effects in well-being indicators such as academic standing or material hardship (Carlson, McLanahan, & England, 2004; Mackenzie, Nicklas, Brooks-Gunn, & Waldfogel, 2011; Wildeman, 2010). Our approach neither aimed to develop new insights into causal processes nor created novel data science techniques to analyze social science data. Rather, we made use of existing methods to thoughtfully navigate the steps required in prediction tasks. Our data after pre-processing and engineering of new features included more than 20,000 features while providing training outcomes for only 2,121 households[2]. Therefore, feature selection was a critical step in our approach[3].

Section 2 of this paper explains our methodology, including pre-processing, engineering, and selection of features in 2.1, and model development in 2.2. Our results are described in Section 3, including model performance (section 3.1) and feature importance (section 3.2). Finally, we close with a discussion (section 4) of insights we obtained from this challenge, and some suggestions for future work related to common prediction tasks in the social sciences.

---

[1] Features are also commonly known as covariates or independent variables.
[2] Out of the total 4,242 households, only 2,121 had training data supplied. The remaining 2,121 were used by the Challenge organizers for leaderboard and hold-out evaluations.
[3] High-dimensional problems where the number of features exceeds the number of observations are not ideal for many machine learning algorithms.

# 2. Methodology

Our team elected to pursue a collaborative approach to the Challenge by dividing the task of generating predictions into three largely independent sub-tasks: preparation of the data, development of models, and aggregation of individual predictions. This modular approach enabled our team members to contribute where their strengths lay to build off each others' work.

We performed a single pre-processing of the data, but used two techniques for feature selection and three distinct learning algorithms. From these three algorithms, four individual prediction sets were generated[4], and four aggregations of these predictions were performed.

## 2.1 Feature Engineering

Most machine learning algorithms are prone to overfitting when their training data contains more features than observations. As this was the case with the raw Challenge dataset, we needed to extract features that could predict the Challenge outcomes, and remove those that would not. Fig. 1 shows how the dataset changed over the course of this study's feature engineering.

**Eliminating features:** We removed any feature that had small variance[5] or contained more than 80% missing data, which reduced the number of features from 12,942 to 5,168.

**Imputation of missing data:** We treated missing data in continuous and ordinal

---
[4] A single learning algorithm (Random Forest) was used by two team members to generate predictions using different data.
[5] Features with absolute variance smaller than 0.05.

features differently from that in categorical features[6].

Since only a small proportion of continuous and ordinal features contained missing values, we performed a simple mean imputation and additionally added two dummy variables[7] when respondents either refused or did not know the answer to a question[8].

One-hot-encoding[9] was performed on the categorical features. Every unique missing code and possible response for a categorical feature became a new dummy variable, such that no imputation was necessary. Our use of one-hot-encoding significantly increased the number of features in our dataset, as each possible response to a categorical question (including every missing code) constituted a new feature. Following this process, the dataset contained 24,864 features for each of the original 4,242 households and no missing data in any of the features[10].

**Composite homelessness features:** Previous research with the FFCWS data has uncovered relationships between features and Challenge outcomes. In one particular study, Fertig et al. (Fertig & Reingold, 2008) identified factors positively and negatively correlated with homelessness or doubling-up (living with someone else). These two sets of features were weighted and aggregated[11] into two composite features that were correspondingly positively and negatively related to homelessness. This resulted in our final, complete, dataset - with 24,866 features for each of 4,242 households.

**Feature selection:** Learning algorithms struggle with high-dimensional data, as was the case at this stage of our methodology, with 6 times as many features (i.e., covariates) as observations (i.e., households). Therefore, we needed to eliminate features that were not predictive of our outcomes. We used two methods to reduce the number of features: (1) univariate feature selection based on mutual information; and (2) extraction of non-zero

---

[6] Our method of identifying features as either continuous or ordinal is found in the supplementary information.
[7] Also identified as binary indicators or boolean variables.
[8] Both of these missing codes (refusal, -1; did not know, -2) could be indicative of an effect present but not tangibly captured by the continuous or ordinal responses in the data.
[9] One-hot-encoding is a process by which features are partitioned into unique response dummy variables. A question with four possible responses (including missing codes) would be replaced with four columns such that the row-wise sum of the resulting variables is exactly one for all observations.
[10] Missing codes are still present as dummy variables created by one-hot encoding.
[11] The exact weights and methodology behind the construction of these features can be found in the supplementary information section of this paper.

LASSO[12] coefficients.

Mutual information (Peng, Hanchuan and Long, Fuhui and Ding, Chris, 2005) is a measure of predictability from information theory defined as:

$$I(X,Y) = \sum_{x \in X} \sum_{y \in Y} p(x,y) \, log(\frac{p(x)p(y)}{p(x,y)}),$$

which captures the level of information that two random variables share[13]. We calculated the mutual information value for each unique outcome (X) and feature (Y) pair. For each outcome we selected the top K[14] features and merged them to create data for distinct K-values that could be used for model building.

LASSO was our second feature selection method (Kukreja, Löfberg, & Brenner, 2006), which admits a penalty parameter (α) that sets coefficients to zero if they are not useful for reducing the model's loss criterion: the sum of squared residuals plus the sum of coefficients' magnitude. Therefore, the LASSO selects features which have predictive power toward the outcome and discards those that do not. The value of α determines the extent of feature selection and was selected such that the resulting regression's $R^2$ (variance accounted for) equaled an ad-hoc value of 0.4 for each outcome. Such a value was large enough to prevent removing too many important features, while still significantly reducing the number of features.

The number of features selected by both methods can be found in the supplementary information section of this paper. It is important to note that feature selection is not directly indicative of feature importance or out-of-sample predictive power. Importance and predictive power are derived from the learning models that are cross-validated, described in the following section.

---

[12] Least Absolute Shrinkage and Selection Operator, or using an L1 norm penalty term in ordinary least squares (OLS) regression to penalize non-zero coefficients.
[13] The mutual information, I(X,Y), is equal to 0 if X and Y are independent as in the case of p(X|Y)=p(X). This means we have no improvement in the knowledge of X from Y. On the other hand, If X and Y are not independent, then I(X,Y)>0: the knowledge of Y is useful to better understand X.
[14] Several values of K were used and can be found in the supplementary information.

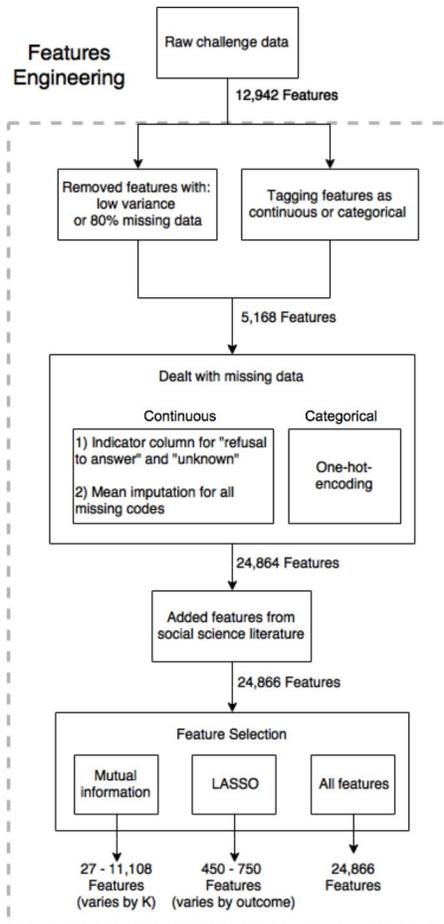

Figure 1: Flowchart of feature engineering with the number of features after every major step of data pre-processing.

## 2.2 Model Building

After feature engineering was completed, we had two databases that could be directly used by learning algorithms to train models, and subsequently generate predictions. In making model design choices, we made use of the leaderboard available to Challenge participants.

Four individual team members developed models in parallel, which resulted in two broad types of approaches: regularized linear models (in the form of an Elastic Net), and non-linear tree-based models (implemented as either Random Forests[15] or Gradient-Boosted Trees).

We treated the prediction of GPA, grit, and material hardship as a continuous regression task, whereas the remaining three outcomes - eviction, job training, and layoff - were predicted as binary, with an underlying probability. For these binary outcomes, we chose to submit the underlying probability of positive class label (1), as opposed to discrete class labels (in this instance, 0 or 1). Predicting probabilities for the binary outcomes would help to improve our performance by lowering the brier loss associated with incorrect predictions[16].

### 2.2.1 The Elastic Net

The Elastic Net is a regularized linear model that combines LASSO (L1) and ridge (L2) regularization[17] (Zou & Hastie, 2005) and achieves the advantages of both methods: sparsity and stability. It can perform additional feature selection by setting coefficients equal to 0, the extent of which is parametrized by the coefficients on the L1 and L2 regularization terms.

In a correctly-specified linear model, the relationship between the independent and dependent variables is linear. The inclusion of only raw un-transformed features could lead to model misspecification and decrease performance. Therefore, we applied three transformations to the continuous features used by the Elastic Net: log, square root, and square, and then normalized each transform-feature pair. The increased number of features did not pose a problem because of the Elastic Net's ability to perform additional feature selection, and in fact, the inclusion of transformed features improved this model's leaderboard performance. Furthermore, we transformed GPA by squaring it, so it exhibited a distribution that was less skewed and closer to normal[18]. Our final model used this GPA transformation, as it improved the model fit when compared to the untransformed performance.

---

[15] The Random Forest algorithm was used by two distinct team members to generate two individual prediction sets.
[16] For instance, for an observation with true value '1' for eviction, if we find that this observation has probability 0.4 of being evicted, we are worse off by predicting 0 (Brier Loss of 1) than by predicting 0.4 (Brier Loss of 0.36).
[17] L1 regularization penalizes proportional to the sum of coefficient magnitudes, and L2 regularization penalizes proportional to the sum of squared coefficient magnitudes.

The Elastic Net generated a single set of predictions for the continuous outcomes only, with regularization parameters selected by k-fold cross-validation. It achieved the best leaderboard results when the continuous features and GPA were transformed and the cutoff for the K-mutual information feature selection method was no more than 300.

## 2.2.2 The Random Forest

The Random Forest algorithm (Liaw & Wiener, 2002) is a non-linear tree-based model and was used by two individual team members. Two unique sets of predictions were generated due to distinct feature selection and validation methods.

One of our team members trained Random Forest regressors or classifiers[19], depending on whether the outcome is continuous or binary. These models were trained on untransformed features selected by mutual information with K = 100[20]. A total of 50[21] Random Forests were trained in a nested cross-validation fashion (Cawley & Talbot, 2010) by generating a series of training/validation/test splits with the given data. Each Forest was fitted to each training split, and its hyperparameters were optimized in the validation splits. Finally, each Forests' predictions were averaged according to performance on the test split. Nested cross-validation can help to prevent Random Forests from overfitting, and this model's final predictions performed well on the binary-valued outcomes of the leaderboard.

---

[18] This transformation of GPA would help prevent problems of model misspecification akin to those for the independent variables.
[19] Regressors predict continuous values, classifiers predict discrete class labels with associated probabilities.
[20] The K-cutoff used was selected based on cross-validation. No significant difference was found with intermediate K values, though extreme values had worse performance in both cross-validation and on the leaderboard.
[21] For the competition, we submitted 200 iterations of a nested Random Forest. However, in this paper we use only 50, which does not require access to high-performance computers, and is reproducible in a reasonable amount of time.

A second team member trained Random Forest regressors on the features selected by the LASSO method. No feature transformations were applied, and the model parameters were selected based on traditional k-fold cross-validation. This individual set of predictions did not perform as well as the other individual predictions on the leaderboard.

### 2.2.3 The Gradient-Boosted Tree

The Gradient-Boosted (GBoost) Tree Model (Friedman, 2001) is a non-linear tree-based method that learns a new decision tree additively to correct the residual errors from the existing sequence of trees. The GBoost Tree is capable of taking into account multiple combinations of features, so we do not have to directly derive combinatorial features manually. Furthermore, the feature sub-sampling function enables us to skip the computationally expensive feature selection step, because the model's training method inherently avoids the overfitting problem.

For this model, we used the imputed 24,864-dimensional training data without feature selection, transformations, or the composite homelessness features we created from social science literature. We used the XGBoost (Chen & Guestrin, 2016; Friedman, 2001)[22] implementation, with XGBRegressor for continuous-valued outcomes and XGBClassifier for binary-valued outcomes. The optimal hyperparameters for GBoost Tree's single set of predictions were selected based on three-fold cross-validation.

### 2.2.4 Ensembled Predictions

Four individual sets of predictions had been generated and submitted to the challenge: one from Elastic Net, two by Random Forest, and another from the GBoost Tree. In an effort to improve generalization, we aggregated our models' predictions in four distinct ways.

---

[22] https://github.com/dmlc/xgboost version 0.6.

First, we performed a simple average of all four predictions, the team average. We averaged all four sets for the continuous outcomes and excluded Elastic Net for the binary ones.

Second, we experimented with a weighted team average, where the weights were determined ad-hoc by relative ranking on the leaderboard. The weight vector for the top three performing predictions for each outcome was given by: [1/2, 1/3, 1/6], for first, second, and third, respectively. Predictions performing worse than 30[th] on the leaderboard were not included in this averaging.

Finally, we looked into aggregation with other models - where we used learning algorithms to find optimal weights for combining our individual prediction sets. This was done in two ways: using either linear/logistic regression or Random Forest regressor/classifier. Cross-validation was performed to select the best hyperparameters for these models.

Our submitted team predictions were generated by the weighted team average, weighted by individual predictions' leaderboard performance.

# 3 Results

We report the performance of all eight prediction sets.

Individual Predictions:

- Elastic Net
- Random Forest with nested cross-validation and mutual information feature selection (Nested RF)
- Random Forest Regressors with LASSO feature selection (LASSO RF)
- XGBoost implementation of Gradient-Boosted Tree

Aggregated Predictions:

- Random Forest aggregation (Ensemble RF)
- Linear regression aggregation (Ensemble LR)
- Weighted Team Average
- Simple Team Average

## 3.1 Model Performance

Model performance for the leaderboard and holdout sets was determined by looking at the improvement over the baseline[23] - or relative accuracy improvement.

The correlation between leaderboard and holdout scores was calculated across outcomes for all models, and for each individual outcome, to assess overfitting to the leaderboard, which was used in developing, evaluating, and aggregating models. The scatterplot of leaderboard vs. holdout performance is shown in Figure 2. Notably, layoff and job training exhibited the largest magnitude correlation coefficients, indicating that performance on the leaderboard was strongly correlated with the performance on the holdout dataset.

The strong correlations present indicate that performance on the leaderboard was a good proxy for performance on the holdout set. That is to say, the leaderboard was the best judge of performance on the holdout set. The same cannot be said for the relation between in-sample error and holdout performance, as we further explore in the supplementary information of this manuscript.

## 3.2 Feature Importance

Feature importance was determined for the Gradient-Boosted Tree, the best-performing of our models. The importance values are derived from the algorithm's ability to partition outcome values depending on feature values. That is, a feature's importance grows as its splits lead to more homogenous subsets of an outcome in subsequent branches. As a result, importance is nearly impossible to interpret for two or more correlated predictors, as they would be equally useful in partitioning outcome values, but the algorithm will only use a single one.

---

[23] The baseline prediction is predicting the simple average value of the training set for the entire sample of households given.

It is important to note that our general approach and usage of machine learning algorithms is not designed to measure causal relationships between features and outcomes. Therefore, the feature importance values for our predictive task should not be confused with the properties we typically associate with parameter estimation tasks. Additional discussion on how to think about these values can be found in (Mullainathan & Spiess, 2017). The top three features for each outcome, along with their importance (as calculated for the GBoost Tree models) and description (as found in the codebook), are provided in Table 1. For the features created through one-hot-encoding, the feature description contains both the value of the response, and the question text. Notably, some of the most important features are closely related to the outcomes, but measured in earlier survey waves.

Values of feature importance were aggregated across categories corresponding to whom the question was posed to, or when the question was asked. This resulted in overall importance of wave (i.e., the year of the data collection) and respondent (e.g., father, mother) in predicting any given outcome. The results of this aggregation are shown in Figure 3. We find that the most important data comes from wave 5 (last wave), except for Material Hardship, and the most important respondent is consistently the mother.

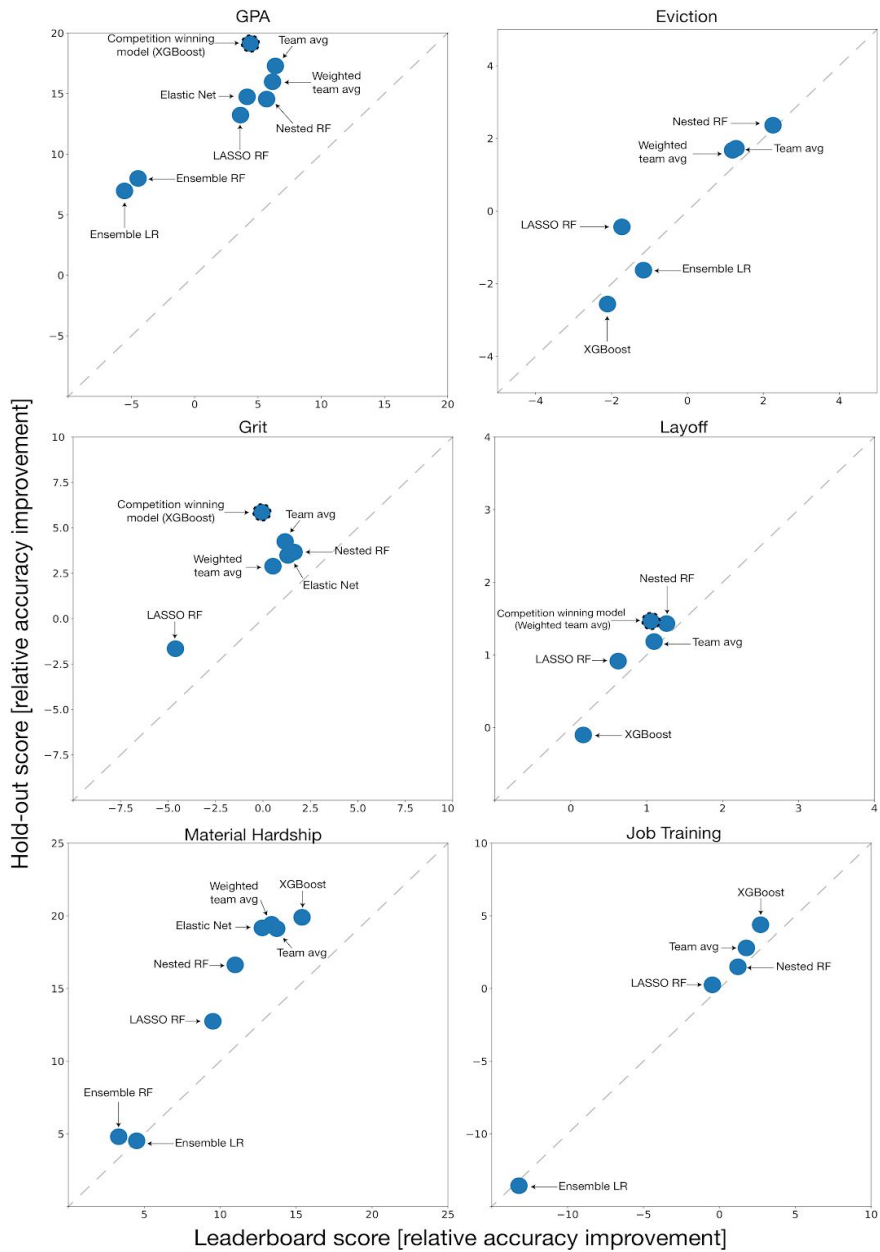

Figure 2: Model performance within the leaderboard and the holdout datasets for each outcome, as relative accuracy improvements over the baseline (average value in the training set). Notable winning and best-performing models are highlighted, and the correlation between leaderboard and holdout scores are calculated overall and for each particular outcome. We have omitted models performing more than 25% worse than the baseline on either leaderboard or holdout sets. Fully labeled model performance on these sets can be found in the supplementary information section.

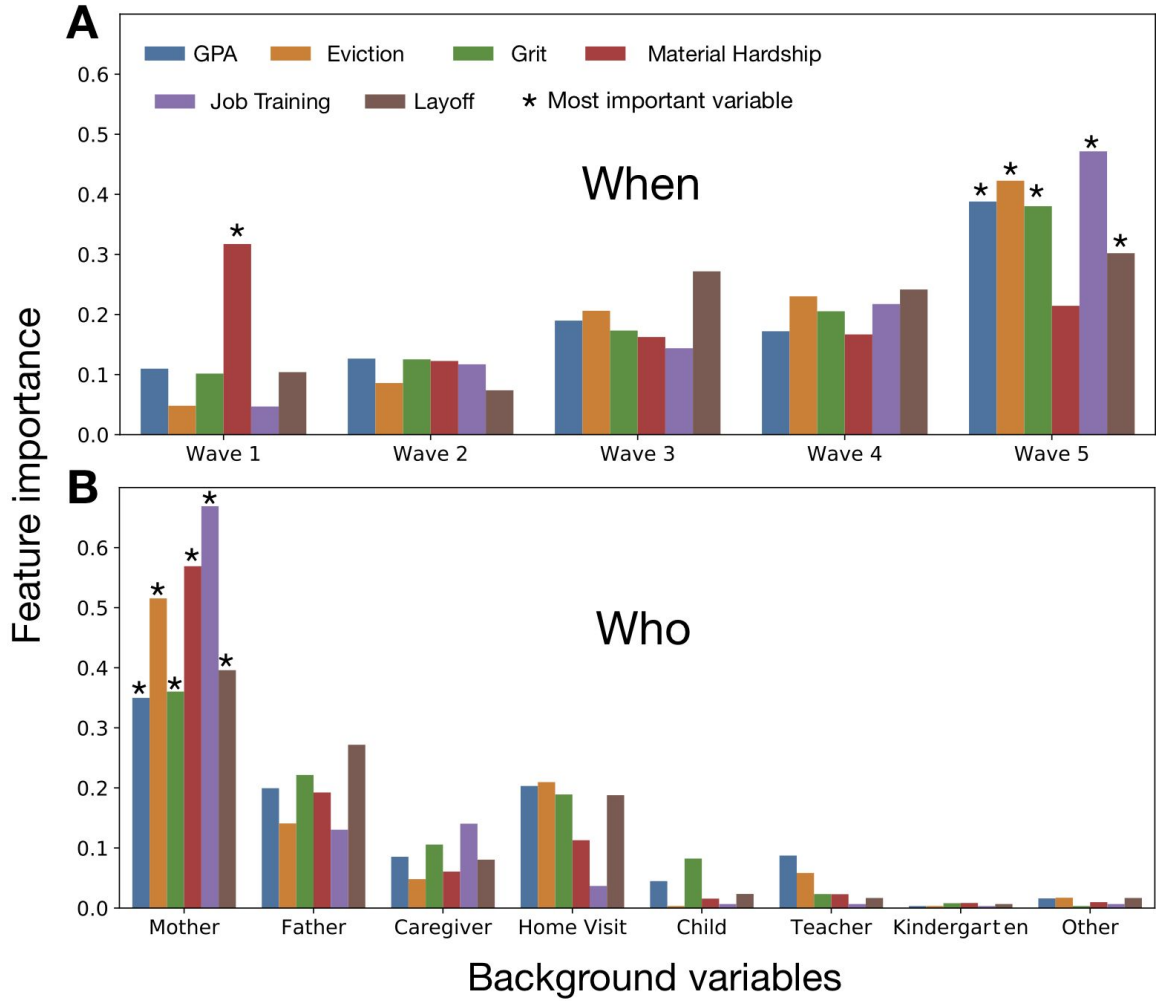

Figure 3: The figure shows the aggregated feature importance of questions asked at particular times (A) or to particular people (B) over the course of the children's lives. These importance values indicate the usefulness of a feature in predicting outcome values, and are neither analogous to coefficients, nor indicate the presence of causal effects. All of these values come from the Gradient-Boosted Tree Model. Stars indicate the highest importance for each outcome.

Table 1: Top-3 most important features for the GBoost Tree Model, per outcome. The feature importance values do not correspond to causal effects.

| Feature Code | Importance | Description |
|---|---|---|
| **GPA** | | |
| hv5_wj10ss | 0.01507 | Woodcock Johnson Test 10 standard score. |
| f3b3 | 0.01004 | How many times have you been apart for a week or more? |
| m2c3j | 0.00904 | How many days a week does father put child to bed? |
| **Grit** | | |
| hv4l47_2 | 0.01520 | Value "2" for: (He/She) stares blankly. |
| hv4r10a_3_1 | 0.01520 | Value "1" for: Any hazardous condition 3: broken glass. |
| hv5_wj9raw | 0.00946 | Woodcock Johnson Test 9 raw score. |
| **M. Hardship** | | |
| m1lenmin | 0.04380 | What was the total length of interview - Minutes. |
| m1citywt | 0.03437 | Mother baseline city weight (20-cities population). |
| m1lenhr | 0.02110 | What was the total length of interview - Hours. |
| **Eviction** | | |
| m5f23k_1 | 0.07216 | Value "Yes" for: Telephone service disconnected because wasn't enough money in past 12 months. |
| m5f23c_1 | 0.05842 | Value "Yes" for: Did not pay full amount of rent/mortgage payments in past 12 months. |
| m3i4 | 0.02062 | How much rent do you pay each month? |
| **Layoff** | | |
| p5j10 | 0.01678 | Amount of money spent eating out in last month. |
| m3i0q | 0.01678 | How important is it: to serve in the military when at war? |
| f5i13 | 0.01678 | How much you earn in that job, before taxes. |
| **Job Training** | | |
| m4k3b_1 | 0.06355 | Value "Yes" for: In the last 2 years, have you taken any classes to improve your job skills? |
| m5i1_1 | 0.06355 | Value "Yes" for: You are currently attending any school/trainings program/classes. |
| m5i3b_1 | 0.06355 | Value "Yes" for: You have taken classes to improve job skills since last interview. |

# 4 Discussion & Conclusion

The best performing model for GPA performed less than 20% better than a simple baseline (i.e., predicting the average GPA for everyone), while the competition-winning grit model had less than 10% improvement over the baseline. We attribute these modest improvements to three main causes.

First, the Challenge data was very high-dimensional, with more features than observations, which was exacerbated by our use of one-hot encoding in the pre-processing step. Furthermore, the traditional machine learning algorithms readily available in software packages were designed for scenarios in which there are more data points than features. Therefore, model performance was extremely sensitive to feature selection. In fact, reruns of an identical model repeatedly resulted in very different leaderboard performance, potentially due to the stochasticity in the algorithms that selected different features and optimal parameters. We believe that high-dimensional scenarios, similar to this Challenge, are becoming more common in computational social science. Such scenarios present a greater need for research and implementation of high-dimensional statistical methods.

Second, common linear models such as ordinary least squares (OLS) and its regularized variations (such as LASSO or Elastic Net) are not ideal for the continuous outcomes in the Challenge, as GPA, grit, and material hardship were bounded. We experimented with Tobit regression (McDonald & Moffitt, 1980) and nonlinear models to address this modeling deficiency; however, Elastic Net still achieved better performance for the continuous outcomes. We believe that bounded regression problems arise in many scenarios and that more attention to developing robust models for bounded regression is warranted. For instance, scikit-learn (Pedregosa et al., 2011), the popular machine learning library in Python, does not currently provide an implementation of a bounded regression such as Tobit (McDonald & Moffitt, 1980).

Third, the de-identification of the data required the omission of information about households' community (e.g., the levels of residential segregation). Previous studies have found that such features can be extremely important for child well-being outcomes.

For example, researchers (Chetty, Hendren, Kline, & Saez, 2014) have found that intergenerational mobility varies substantially across geographic areas. This study found that community-level features (e.g., residential segregation, income inequality, family stability, and social capital) were the most predictive of intergenerational mobility ($R^2 = 0.38$). Perhaps a second and more secure stage of the challenge that allowed access to geographical or pre-computed community indicators would allow models to perform better and provide insight as to how location-variant features may affect the outcomes of children's lives, while preserving the privacy of households.

As illustrated by the Challenge organizers, most submitted models captured a very small portion of the variance in the outcomes; with even the best models predicting around the average. We believe these observations indicate poor predictive performance. In addition to the technical reasons above, we speculate that the inherent unpredictability of this setting could serve as a more fundamental reason behind the poor performance of models. This hypothesis becomes more plausible in light of recent focus on the limits to prediction and purely random outcomes, analogous to luck, in complex social settings such as ours (Hofman, Sharma, & Watts, 2017).

We believe that constructing predictive features from raw features may have contributed to our high relative performance. Fortunately, there is a vast body of research knowledge, not just restricted to Fragile Families data but in other similar contexts, that has studied the causal factors that affect the well-being of children. The inclusion of this knowledge in models such as ours could significantly affect predictive performance and improve the ability to verify previously published findings. However, as we experienced, a manual review of such a vast body of knowledge is next to impossible for researchers who lack domain knowledge or expertise in the sociology of Fragile Families. For those who participate without extensive domain expertise, we believe the existence of a database incorporating the main results of relevant social science studies in a queryable structure should greatly help performance in prediction tasks - not only for the Challenge but for evaluating the effectiveness of interventions in many other problem domains important to policymaking.

# Supplementary Information for: Winning Models for GPA, Grit, and Layoff in the Fragile Families Challenge

## Code Repository

The code used to create the predictions analyzed in this manuscript can be found online at: https://github.com/drigobon/FFC_Pentlandians_Code. This repository does not include any of the data obtained from the FFCWS that was used in this study.

Furthermore, we note that the predictions submitted to the Challenge are not identical to those used in this study. The results generated from the above repository are slightly distinct, and less computationally intense, in order to favor reproducibility of our analysis. Specifically, the number of models used to generate the Nested Random Forest has been decreased so our code will run in a reasonable amount of time without high-performance computers.

## Ordinal or Continuous Feature Criteria

In our data pre-processing steps, we treated categorical features differently from features which indicated ordered responses (i.e. continuous or ordinal). To identify ordinal and continuous features among the given data, we used the provided list of question metadata and a combination of two heuristics: i) features with more than 15 unique values; or ii) descriptions containing keywords such as "How many," "Rate," "Frequency" or "Total," would be most likely to have continuous or ordinal responses. In the identification of these features, we specifically looked for keywords that would indicate the presence of an 'order' in answers given. The exact logical keyword heuristic is shown here:

(*How* & (*Is* | *Many* | *Often* | *Much* | *Long*)) | *Rate* | *Frequency* | *Number* | *#* | *Level* | *Highest* | *Amount* | *Days* | *Total* | *Scale* | *Times.*

However, not all of the continuous and ordinal features were properly identified by these two criteria. Thus, we conducted a manual review and correction of the question text, which ultimately resulted in 3,682 of the 5,168 original features identified as categorical, while the remaining 1,486 contained ordered responses. With this information, we could proceed with mean imputation of the continuous and ordinal features, and one-hot-encoding for the categorical features.

# Feature Selection

We used two distinct methods of selecting features for model building: mutual information and LASSO. With over 24,000 features, most learning algorithms would overfit to our training data. This section details the number of features selected by our two methods and their relative similarity in selected features.

## Mutual Information Feature Selection

Mutual information was our first method of feature selection, where we selected the top K features in mutual information with each outcome. We selected the top K for each individual outcome and merged them to create data matrices to use for the model building. Table S1 below shows the number of features selected at each K-value used in this study. Notably, some of the features overlap with as small a K-value as 5.

Table S1: Number of features selected by the mutual information cutoff for various values of K per outcome.

| K-Value | Number of Total Features Selected |
| --- | --- |
| 5 | 27 |
| 15 | 83 |
| 50 | 261 |
| 100 | 498 |
| 200 | 941 |
| 300 | 1417 |
| 500 | 2224 |
| 700 | 2914 |
| 1000 | 3965 |
| 1500 | 5549 |
| 2000 | 7010 |
| 3000 | 9377 |
| 4000 | 11255 |

## LASSO Feature Selection

The LASSO feature selection method selects features with non-zero coefficients for regressions run on each individual outcome variable. This is effective because of the L1 regularization parameter, $α$, that penalizes the sum of the coefficients and results in many of them equal to 0. The regularization parameter was selected so that the $r^2$ value of the regression (variance accounted for) was as close to 0.4 as possible. The value of $α$, the $r^2$ value, and the number of features selected by this method for each outcome can be found in Table S2. We note that the granularity of the $α$ parameter was not sufficiently small to have an $r^2$ value very close to the target of 0.4.

Table S2: Number of features, $r^2$ value, and regularization parameter for the LASSO feature selection method in each outcome.

| | Number of Features | $r^2$ Value | α |
|---|---|---|---|
| GPA | 314 | 0.344 | 0.129 |
| Grit | 86 | 0.374 | 0.021 |
| Material Hardship | 353 | 0.278 | 0.021 |
| Layoff | 527 | 0.343 | 0.021 |
| Eviction | 65 | 0.264 | 0.021 |
| Job Training | 77 | 0.326 | 0.021 |

## Feature Selection Comparison: Mutual Information vs LASSO

In order to study the effectiveness of the feature selection methods used in this study, we compared the features selected by both mutual information and LASSO at various cutoffs. Specifically, we looked at K ∈ {5, 15, 50, 100, 200, 300, 500, 700, 1000, 1500,

2000, 3000, 4000}, and at $r^2 \in$ {0.1, 0.2, 0.3, 0.4, 0.5, 0.6, 0.7, 0.8, 0.9}. The value of the heatmap shown in Fig. S1 indicates the intersection over the union of both methods (also known as the Jaccard coefficient), that is, the number of features selected by both methods over the total number of features selected by either. There is little similarity between the resulting features, with a maximum of 0.13 for the least stringent cutoffs for mutual information and LASSO.

However, the aforementioned analysis does not account for the fact that highly collinear features may be present in data from both feature selection methods. In order to study this, we extracted the first fitted principal component of data matrices selected by both mutual information and LASSO at various K and $r^2$ cutoffs, respectively. The correlation coefficient between these components was calculated and plotted on the heatmap shown in Fig. S2. Notably, the first principal component of the LASSO-selected variables is particularly invariant to the $r^2$ cutoff selected, and beyond K = 20 there is very little change in the correlation coefficient.

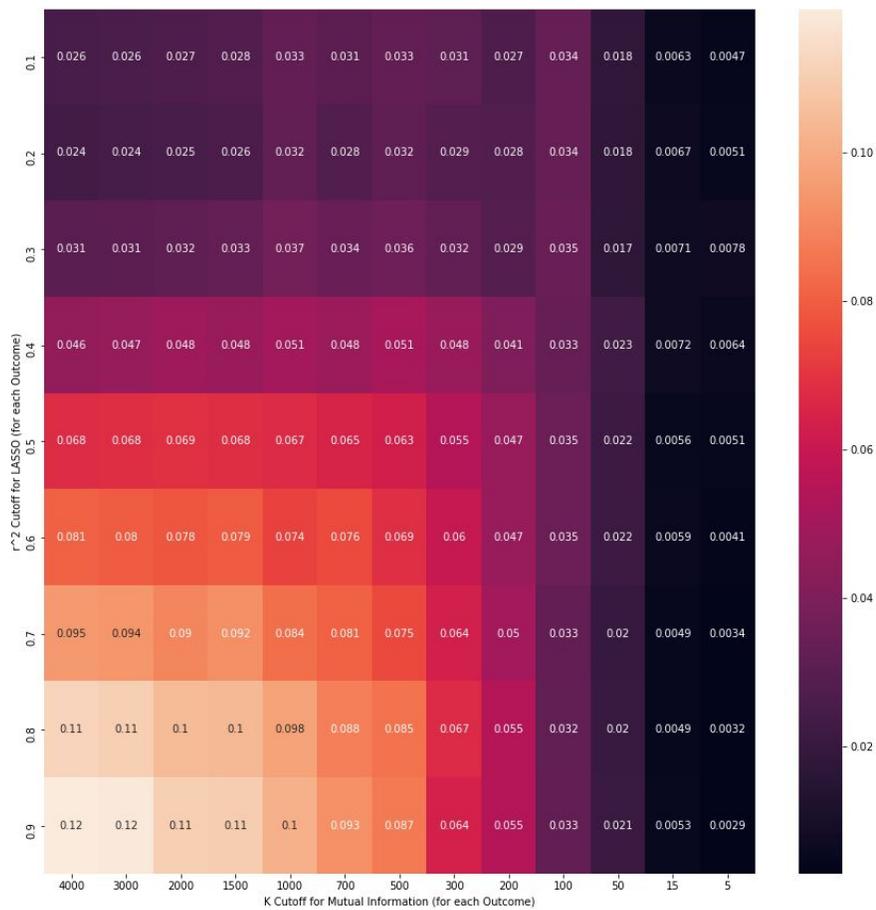

Figure S1: Comparison of features selected by Mutual Information or LASSO at various cutoffs for K or $r^2$, respectively. The value shown in the heatmap is a proportion, calculated as the number of elements in the union of the features selected divided by the number of elements in the intersection. It indicates how many features were selected by both over the number of features selected by either (also known as the Jaccard coefficient).

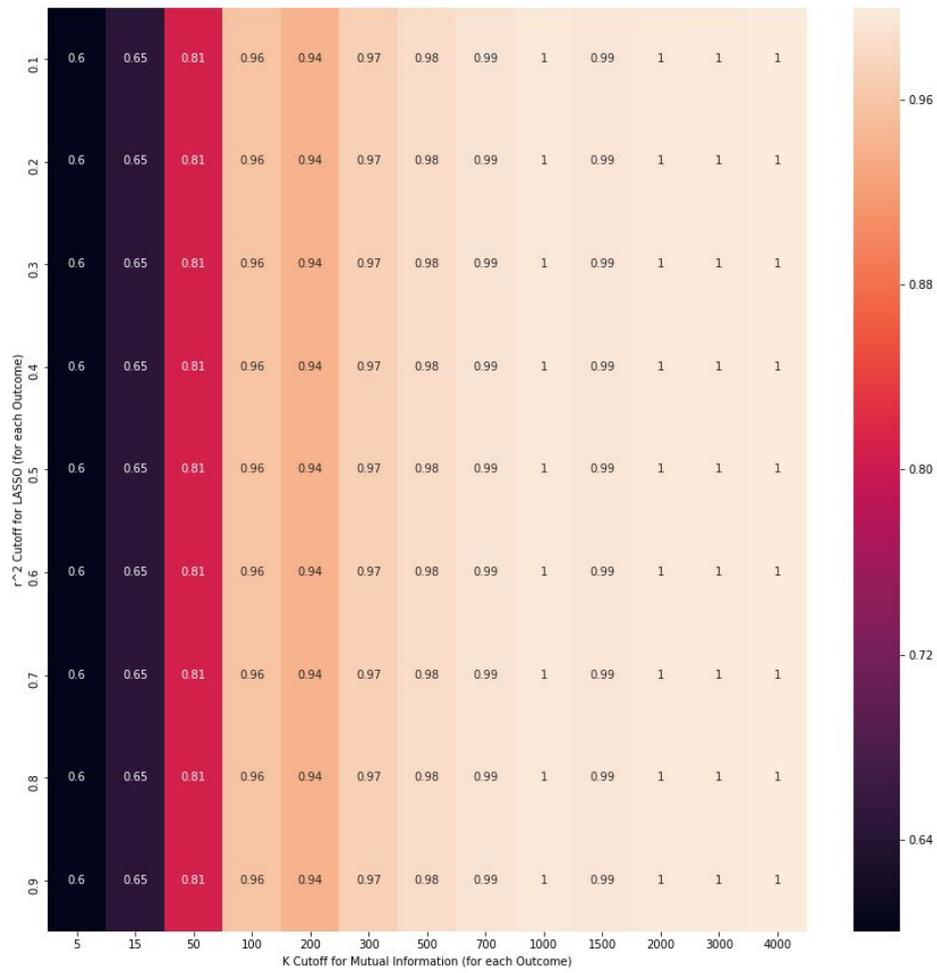

Fig. S2: Correlation coefficient calculated for the first fitted principal component of matrices resulting from a particular $r^2$ or K cutoff for LASSO or mutual information feature selection methods, respectively. These values indicate the similarity of features selected by both methods, accounting for collinearity in these features.

# Construction of Composite Homelessness Features from External Literature

Two composite features were created in this study, both of which were based on previous research using the Fragile Families Dataset. These features were identified as being positively correlated with homelessness, and negatively correlated with homelessness.

The first feature was created from a simple sum of the features: 1) mother receives welfare, 2) mother resides in public housing, 3) mother lives with father, 4) mother's race, and 5) number of children. We defined 'mother's race' as 3 only if the mother was either black or Hispanic, 0 otherwise, and the number of children was capped at 3.

The second was created from the sum of the following features: 1) mother family or friends willingness to help, 2) mother has lived in the neighborhood more than 5 years, and 3) the number of moves in the first year after birth.

Many of the questions used to create our two composite features were asked multiple times over the course of the study. In aggregating all responses to an identical question posed at different waves, we selected to weight the most recent (wave 5) response 3x more than previous ones.

# Holdout, Leaderboard, and In-Sample Results

Model performance in the leaderboard and holdout sets was calculated by relative performance improvement over the baseline. In this study, the baseline was defined by the average training value for each outcome. The performance of all models is reported in Figure S3.

In making modeling choices, we made use of the Challenge leaderboard as out-of-sample feedback. The leaderboard motivated choices such as: what K-value cutoff for mutual information to use, transformations for the Elastic Net, and ensembling with our weighted team average.

We strongly believe that our use of the leaderboard helped us expand our available training data. Fig. 2 in the main text highlights the strong correlation between Leaderboard and Holdout scores. However, the same cannot be said for in-sample improvement over the baseline. This section contains two plots, Figs. S4 and S5, that show individual correlations by outcome between in-sample and leaderboard, holdout, respectively. We notice that there is no strong relationship here, certainly weaker than that shown in Fig. 2 between the Leaderboard and the Holdout sets. The lack of a significant relationship in Figures S4, S5 may be an indication of overfitting to the training set, as the performance improvement observed in-sample does not generalize to either of the out-of-sample evaluation datasets.

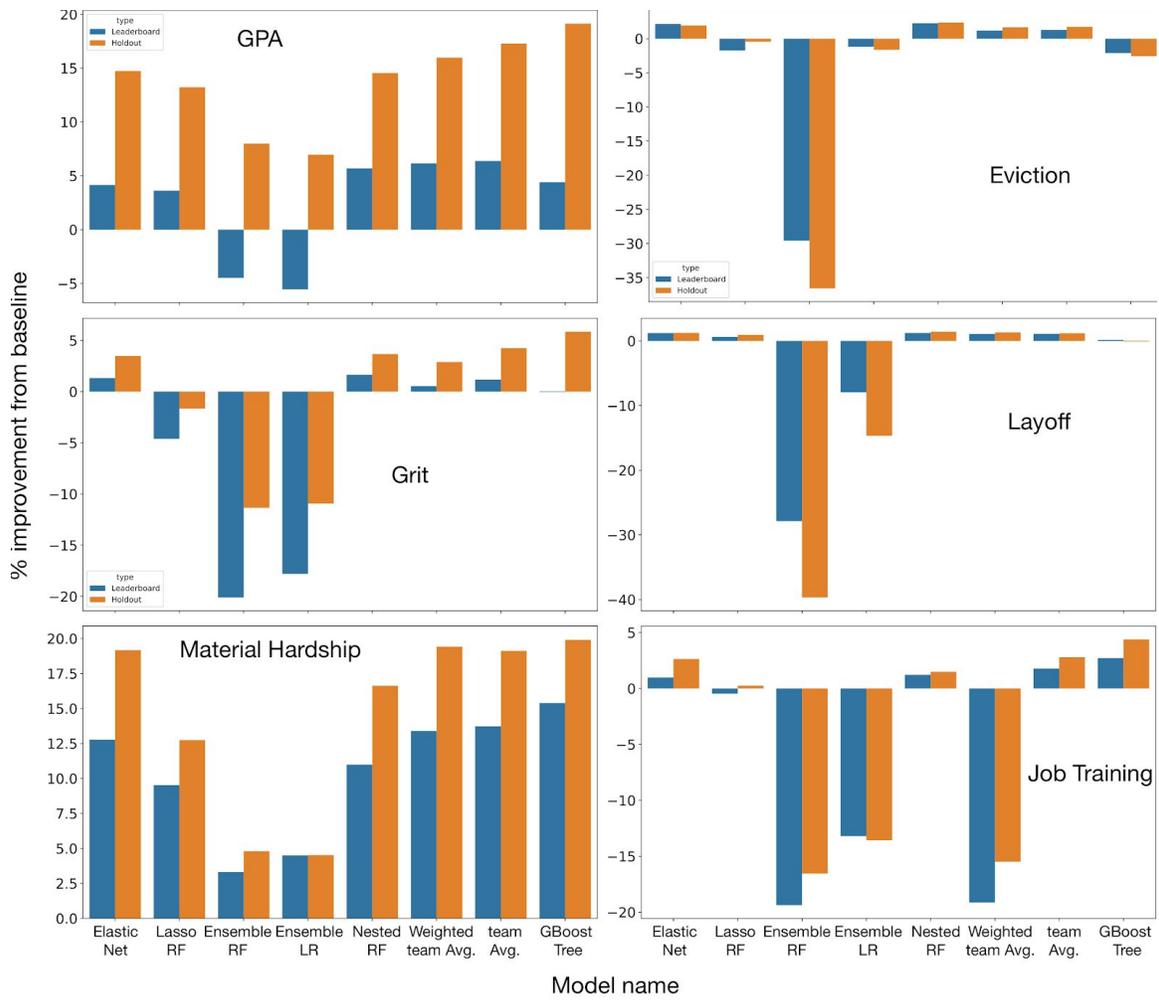

Figure S3: Performance of all prediction sets in the leaderboard and holdout sets. Performance is measured by percentage improvement over the baseline or relative accuracy improvement.

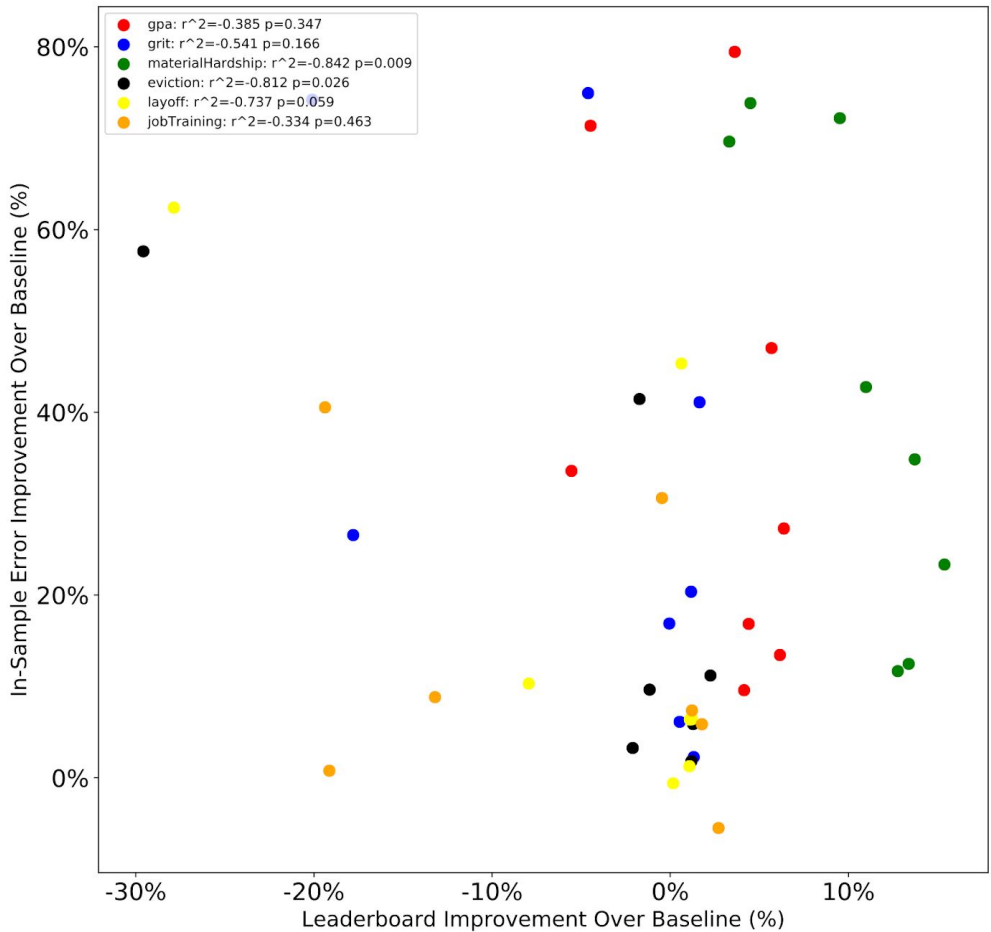

Figure S4: Scatterplot of all submitted models, showing both an improvement over the baseline for Leaderboard and In-Sample. The baseline was defined by the average value in the training set for each outcome. Correlations per outcome can be found in the legend.

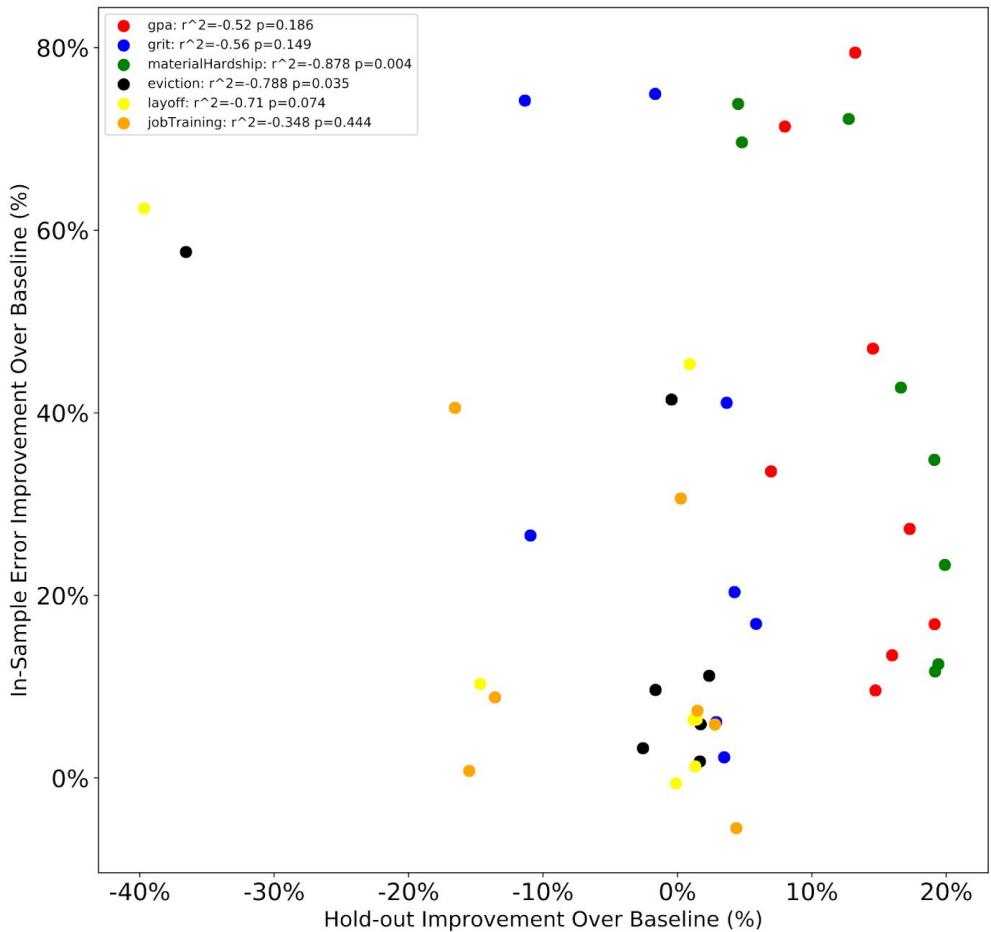

Figure S5: Scatterplot of all submitted models, showing both improvements over the baseline for Holdout and In-Sample. The baseline was defined by the average value in the training set for each outcome. Correlations per outcome can be found in the legend.

# Model Hyperparameters and Feature Importance

In this section, we elect to focus on two models which represent our broad types of models: regularized linear (Elastic Net), and non-linear tree-based (Gradient-Boosted Tree). For both of these models, we report the optimal hyperparameters selected by cross-validation - in the case of Elastic Net, we only report these values for the continuous outcomes, as it did not predict any of the binary outcomes. We also provide information regarding the importance of features in the Gradient-Boosted Tree.

## Elastic Net

The Elastic Net exhibited strong performance for the continuous outcomes on both the leaderboard and the holdout set. Table S3 details the hyperparameters selected by cross-validation for a single prediction set, along with the resulting number of non-zero coefficients in the model. These hyperparameters resulted in 58 non-zero coefficients for GPA, 33 for grit, and 58 for material hardship. It is important to note that there is significant stochasticity in the selection of these parameters, and further training of the same data without setting the random seed may yield distinct optimal values.

Table S3: Optimal parameters and resulting non-zero coefficients for the Elastic Net's prediction of continuous outcomes. These parameters are stochastic, and selected through cross-validation

| Parameter | GPA | Grit | Material Hardship |
|---|---|---|---|
| alpha | 0.1163 | 0.0164 | 0.0022 |
| l1_ratio | 0.5 | 0.5 | 0.99 |

## Gradient-Boosted Tree

The Gradient-Boosted Tree was the overall best-performing individual model detailed in this manuscript. Table S4 shows the optimal hyperparameters for each outcome, as selected by grid-search cross-validation. These hyperparameters are stochastic, and may not be identical if the cross-validation were to be run again. Tables S5-10 indicate the top 10 most important features in predicting the outcomes, ordered top-down and including both variable name, importance, and description. The descriptions and associated values (for the one-hot-encoded features) come directly from the codebook metadata given to Challenge participants. Notably, some of the most important features are equivalent to the outcomes but measured in earlier survey waves.

The feature importance for the Gradient-Boosted Tree is a 'score' indicating how useful a given feature was in constructing decision trees within the model. The score is formally calculated by the sum of gini-impurity (a measure of how often a randomly chosen element from the set would be incorrectly labeled if it was randomly labeled according to the distribution of labels in the subset) gain of a feature in all trees. Generally speaking, if a feature is consistently used to split samples, it will have a higher importance. These importance values are not to be interpreted as coefficients.

Figure S6 indicates the structure of an individual Gradient-Boosted Tree regressor along with a short explanation of its method, and is part of the actual Gradient-Boosted Tree model used for the competition. It is difficult to visualize all the model's unique decision trees, therefore the figure is not comprehensive.

Table S4: Optimal parameters selected based on grid-search cross-validation[24].

| Parameter | GPA | Grit | Material Hardship | Eviction | Layoff | Job Training |
|---|---|---|---|---|---|---|
| colsample_bytree | 0.4 | 0.8 | 0.8 | 0.6 | 0.8 | 0.4 |
| learning_rate | 0.01 | 0.01 | 0.01 | 0.02 | 0.05 | 0.02 |
| max_depth | 2 | 2 | 5 | 2 | 2 | 2 |
| n_estimators | 1000 | 1000 | 1000 | 100 | 100 | 100 |
| subsample | 0.6 | 0.6 | 0.4 | 0.6 | 0.6 | 0.8 |

---

[24] Grid-search cross validation is an exhaustive search on the discretized parameter grid.

Table S5: Top-10 Feature Importance Codes and Descriptions for GPA. These importance values are not analogous to coefficients, and do not imply causal effects.

| Variable Name | Importance | Description |
| --- | --- | --- |
| hv5_wj10ss | 0.01507 | Woodcock Johnson Test 10 standard score. |
| f3b3 | 0.01004 | How many times have you been apart for a week or more? |
| m2c3j | 0.00904 | How many days a week does father put child to bed? |
| m1i1 | 0.00904 | What is the highest grade/years of school that you have completed? |
| hv4k2_expen | 0.00803 | Total expense for food used at home. |
| hv5_ppvtss | 0.00737 | PPVT standard score. |
| hv5_wj9ss | 0.00703 | Woodcock Johnson Test 9 standard score. |
| m5d20 | 0.00636 | First principal component scale created from m5d20a-p. |
| m2h18c_3 | 0.00636 | Value "No" for: Did you fill out a federal tax return for previous full year? |
| cm2povco | 0.00603 | Constructed - Poverty ratio - mother's household income/poverty threshold. |

Table S6: Top-10 Feature Importance Codes and Descriptions for Grit. These importance values are not analogous to coefficients, and do not imply causal effects.

| Variable Name | Importance | Description |
| --- | --- | --- |
| hv4l47_2 | 0.01520 | Value "2" for: (He/She) stares blankly. |
| hv4r10a_3_1 | 0.01520 | Value "1" for: Any hazardous condition 3: broken glass. |
| hv5_wj9raw | 0.00946 | Woodcock Johnson Test 9 raw score. |
| k5g1b_3 | 0.00878 | Value "Always" for: Even when a task is difficult, I want to solve it anyway. |
| cf2b_age | 0.00844 | Constructed - Baby's age at time of father's one-year interview (months). |
| m5c6 | 0.00743 | He is fair and willing to compromise when you have a disagreement. |
| hv5_ppvtss | 0.00709 | PPVT standard score. |
| cm1hhinc | 0.00642 | Constructed - Household income (with imputed values). |
| p5i26 | 0.00642 | Frequency you know what child does during free time. |
| k5g2h_0 | 0.00608 | Value: "Not at all true" for: It's hard for me to finish my schoolwork. |

Table S7: Top-10 Feature Importance Codes and Descriptions for Material Hardship. These importance values are not analogous to coefficients, and do not imply causal effects.

| Variable Name | Importance | Description |
|---|---|---|
| m1lenmin | 0.04380 | What was the total length of interview - Minutes. |
| m1citywt | 0.03437 | Mother baseline city weight (20-cities population). |
| m1lenhr | 0.02110 | What was the total length of interview - Hours. |
| cm1age | 0.01609 | Mother's age (years). |
| m1a12a | 0.01217 | How many other biological children do you have? |
| m1b1a | 0.01053 | How many years did you know Baby's Father before you got pregnant? |
| m1e1d1 | 0.00770 | People who currently live in your HH - 1st age? |
| m1e1d2 | 0.00524 | People who currently live in your HH - 2nd age? |
| m1f1a | 0.00479 | How long have you lived in neighborhood - Years? |
| m1b12a | 0.00433 | In last mo, how often did you and BF disagree about money? |

Table S8: Top-10 Feature Importance Codes and Descriptions for Eviction. These importance values are not analogous to coefficients, and do not imply causal effects.

| Variable Name | Importance | Description |
|---|---|---|
| m5f23k_1 | 0.07216 | Value "Yes" for: Telephone service disconnected because wasn't enough money in past 12 months. |
| m5f23c_1 | 0.05842 | Value "Yes" for: Did not pay full amount of rent/mortgage payments in past 12 months. |
| m3i4 | 0.02062 | How much rent do you pay each month? |
| m5f23c_2 | 0.02062 | Value "No" for: Did not pay full amount of rent/mortgage payments in past 12 months. |
| m5f23k_2 | 0.02062 | Value "No" for: Telephone service disconnected because wasn't enough money in past 12 months. |
| m5i3c_1 | 0.02062 | Value "Yes" for: You received any kind of employment counseling since last interview. |
| m5f23a_1 | 0.01718 | Value "Yes" for: Received free food or meals in past 12 months. |
| f1citywt_rep1 | 0.01718 | Father baseline city replicate weight no. 1. |
| m3d9 | 0.01718 | Last month of relationship with father. |
| f4i4 | 0.01718 | How much rent do you pay each month? |

Table S9: Top-10 Feature Importance Codes and Descriptions for Layoff. These importance values are not analogous to coefficients, and do not imply causal effects.

| Variable Name | Importance | Description |
|---|---|---|
| p5j10 | 0.01678 | Amount of money spent eating out in last month. |
| m3i0q | 0.01678 | How important is it: to serve in the military when at war? |
| f5i13 | 0.01678 | How much you earn in that job, before taxes. |
| f4i23m_2 | 0.01678 | Value "No" for: In past 12 months, you worked overtime or taken a second job? |
| hv3b7_3_1 | 0.01678 | Value "1" for: Part of bedtime routine -- change diaper/take to toilet? |
| f3k22 | 0.01342 | In last year, how many wks. did you work all regular jobs? |
| m4f2d2_6 | 0.01342 | Value "6" for: What is second person's relationship to you? |
| m5f7a_2 | 0.01342 | Value "No" for: Received help from an employment office in past 12 months. |
| m3i23d_2 | 0.01007 | Value "No" for: In past year, did you not pay full gas/oil/electricity bill? |
| p5j2e | 0.01007 | Number of servings of regular soda child has on typical day. |

Table S10: Top-10 Feature Importance Codes and Descriptions for Job Training. These importance values are not analogous to coefficients, and do not imply causal effects.

| Variable Name | Importance | Description |
|---|---|---|
| m4k3b_1 | 0.06355 | Value "Yes" for: In the last 2 years, have you taken any classes to improve your job skills? |
| m5i1_1 | 0.06355 | Value "Yes" for: You are currently attending any school/trainings program/classes. |
| m5i3b_1 | 0.06355 | Value "Yes" for: You have taken classes to improve job skills since last interview. |
| p5l13f_1 | 0.05017 | Value "Yes" for: Gifted and talented program. |
| p5l13f_2 | 0.04013 | Value "No" for: Gifted and talented program. |
| m5i19a | 0.03010 | Amount earned from all regular jobs in past 12 months. |
| m4l2_1 | 0.02676 | Value "Yes" for: In past 12 months have you given/loaned any money to friends or relatives? |

| cm5edu_3 | 0.02007 | Value "Some coll, tech" for: Mother's education: year 9. |
| m5i3b_2 | 0.01672 | Value "No" for: You have taken classes to improve job skills since last interview. |
| cf5hhinc | 0.01338 | Constructed - Father's Household income (with imputed values). |

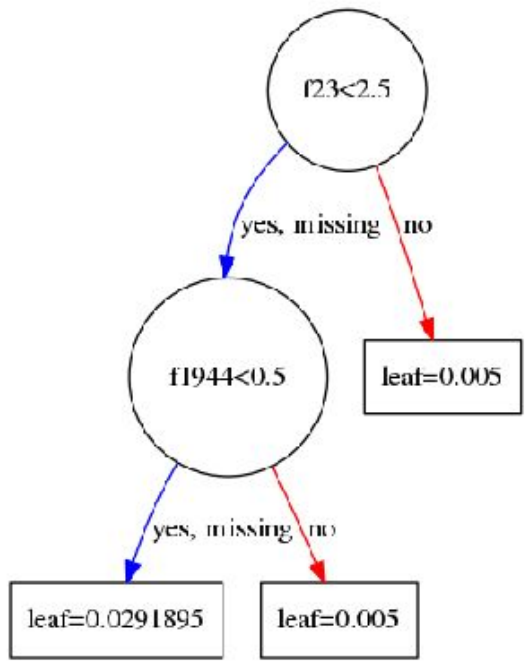

Figure S6: Example of an individual tree of an XGBoost regressor (implementation of Gradient-Boosted Tree). This is a single decision tree in the ensemble generated by the XGBoost implementation of this learning algorithm. In this example, the tree has two features (f23 and f1994) to branch an input sample into one of the three leaf nodes which have scored. The score of the leaf node that a sample is classified will be added to the final prediction value of the model.